\documentclass[aps,pre,showpacs,amssymb,a4paper,twocolumn]{revtex4-1}%
\usepackage{amsmath}%
\usepackage{amsfonts}%
\usepackage{amssymb}%
\usepackage{graphicx}

\begin{document}

\title{Probing ergodicity in granular matter}
\author{Fabien Paillusson}
\author{Daan Frenkel}
\affiliation{Department of Chemistry, University of Cambridge, Lensfield Road, Cambridge, CB2 1EW, U.K.}
\date{\today}
\begin{abstract}
When a granular system is tapped, its volume changes. Here, using a
well-defined macroscopic protocol, we prepare an ensemble of granular
systems and track the statistics of volume changes as a function of the
number of taps. This is in contrast to previous studies, which have focused
on single trajectories and assumed ergodicity.  We devise a new method to
assess the convergence properties of a sequence of ensemble volume
histograms and introduce a reasonable approximate version of an invariant
histogram. We then compare these invariant histograms with histograms
generated by sampling a long trajectory for one system and observe
nonergodicity, which we quantify. Finally, we use the overlapping histogram
method to assess potential compatibility with Edwards' canonical
assumption. Our histograms are incompatible with this assumption.
\end{abstract}
\maketitle

\section{Introduction}
Controlling the packing density of a vibrated granular
system is of industrial and scientific interest. In this paper, we focus on
sequences of packings that are in mechanical equilibrium. For such packings,
the effect of shaking depends solely on the tapping stage of the shaking
protocol. This stage is uniquely parameterized by the extra momentum given
to the system that sets the {\it amount of dilation} allowed after each tap
\cite{Mueggenburg12}.

The protocol used to prepare the granular system is also of primary
importance. Currently, no macroscopic preparation protocol exists that
allows one to choose the initial microscopic configuration exactly. Due to
the inevitable dispersion in the initial conditions, the only reproducible
quantity that one can measure in experiments is the probability of observing
a given volume after a given number of taps.

During the last twenty years, properties of the time averages of packing
volumes and their fluctuations have been studied extensively
\cite{PicaCiamarra07,
  Nowak98,Yu06,Pugnaloni08,Brey01,Mueggenburg12}. Recently McNamara et
al. \cite{McNamara09} carried out a careful analysis of volume histograms
sampled over time. However, if the systems under study are not ergodic---an
{\it a priori} assumption in most of these studies---then time averages over fluctuations
for one system need not be the same as averages over the initial conditions. To tackle this problem, one should ideally
study an ensemble of systems prepared in the same macroscopic way and follow
the evolution of their volume with time. To our knowledge, such a study is
still lacking.

In this paper, we use numerical simulations to study the evolution of an
ensemble of granular systems prepared with the same macroscopic protocol. We
analyse the rate of convergence of these histograms and find that, under
certain conditions, individual trajectories are nonergodic.  Finally, we
find that our results are incompatible with Edwards' hypothesis, namely,
that the dependence of the volume statistics on preparation protocol can be
captured by a simple Boltzmann weight \cite{Edwards89}.

\section{``Equilibrium'' distribution}
\subsection{Numerical simulation}
The shaking experiment is simulated using a dissipative, event-driven MD
simulation of frictionless hard spheres subject to gravity. We use lateral
periodic boundary conditions and place a wall at the bottom of the
simulation box.

A known problem with event-driven schemes applied to dissipative systems is
that a linear restitution coefficient makes the simulation ``freeze''
locally in higher-density regions \cite{Goldhirsch93}. To avoid this problem,
we use a nonlinear restitution coefficient, so that the spheres do not loose
all their kinetic energy, even after many collisions. When the system
relaxes after an external perturbation (e.g. a single tap), the final state
will have a well-defined structure, but the spheres will still move
locally---albeit with a typical displacement amplitude that is much smaller
than their size. We assume that the packings thus obtained are
representative of the typical stable configurations that are obtained in
real experiments at sufficiently high densities.

We start every simulation by generating an equilibrated fluid configuration
of 1000 hard spheres at packing fraction $\phi = 0.2$ in a cubic simulation
box. We then switch on gravity and let the system reach mechanical
equilibrium, as explained above. This is the preparation stage of the
simulation.  Once the packing has settled, we ``tap'' the system by
accelerating the bottom layer of the spheres impulsively.  Concretely, we
add to these spheres the vertical velocity $A \sqrt{ga}$, where $A$ is
a dimensionless tapping amplitude, $g$ is the acceleration of gravity, and
$a$ is the diameter of the spheres.  We repeat this process at regular time
intervals, which are longer than the typical time it takes the
packing to settle after a perturbation.

\subsection{Volume histogram evolution}
To study the ensemble statistics of our system, we prepare about 50000
stable initial conditions as described above. The corresponding distribution
of the initial volumes is denoted $\rho_0(V)$. For a given amplitude $A$,
the volume histogram after $i$ tapping steps is denoted $\rho^{^A}_i(V)$.
The volume of a packing is defined as the smallest axis-aligned cuboidal
volume that completely contains all the spheres, and whose bottom face
lies in the plane of the wall.

In analogy with the Gibbs approach in statistical mechanics, we define the
statistical ``equilibrium'' ensemble distribution as the asymptotic volume
distribution obtained when the number of taps tends to infinity. If there is
a well-defined time-scale for approach to the steady state, then this
procedure should yield a good approximation of the invariant distribution.

\begin{figure}[h!]
    \includegraphics[scale=0.6,keepaspectratio=true]{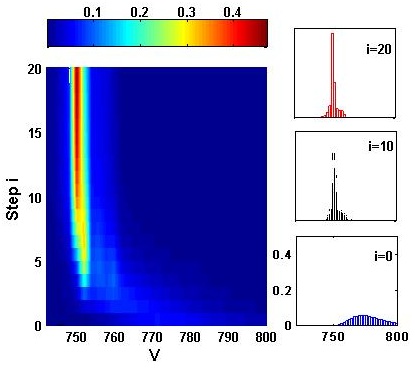}
\caption{Typical volume histogram evolution for an ensemble of about 50000
  replicas of a system of 1000 spheres. The picture shown corresponds to an
  amplitude $A$, equal to 6.  }
\label{Fig1}
\end{figure}

In Fig.\ref{Fig1} we show the evolution of a typical volume histogram as a
function of the number of tapping steps.  Starting from a very broad volume
distribution, the ensemble histograms become narrower, and eventually reach
an invariant shape.

\begin{figure}[h!]
    \includegraphics[scale=0.2,keepaspectratio=true]{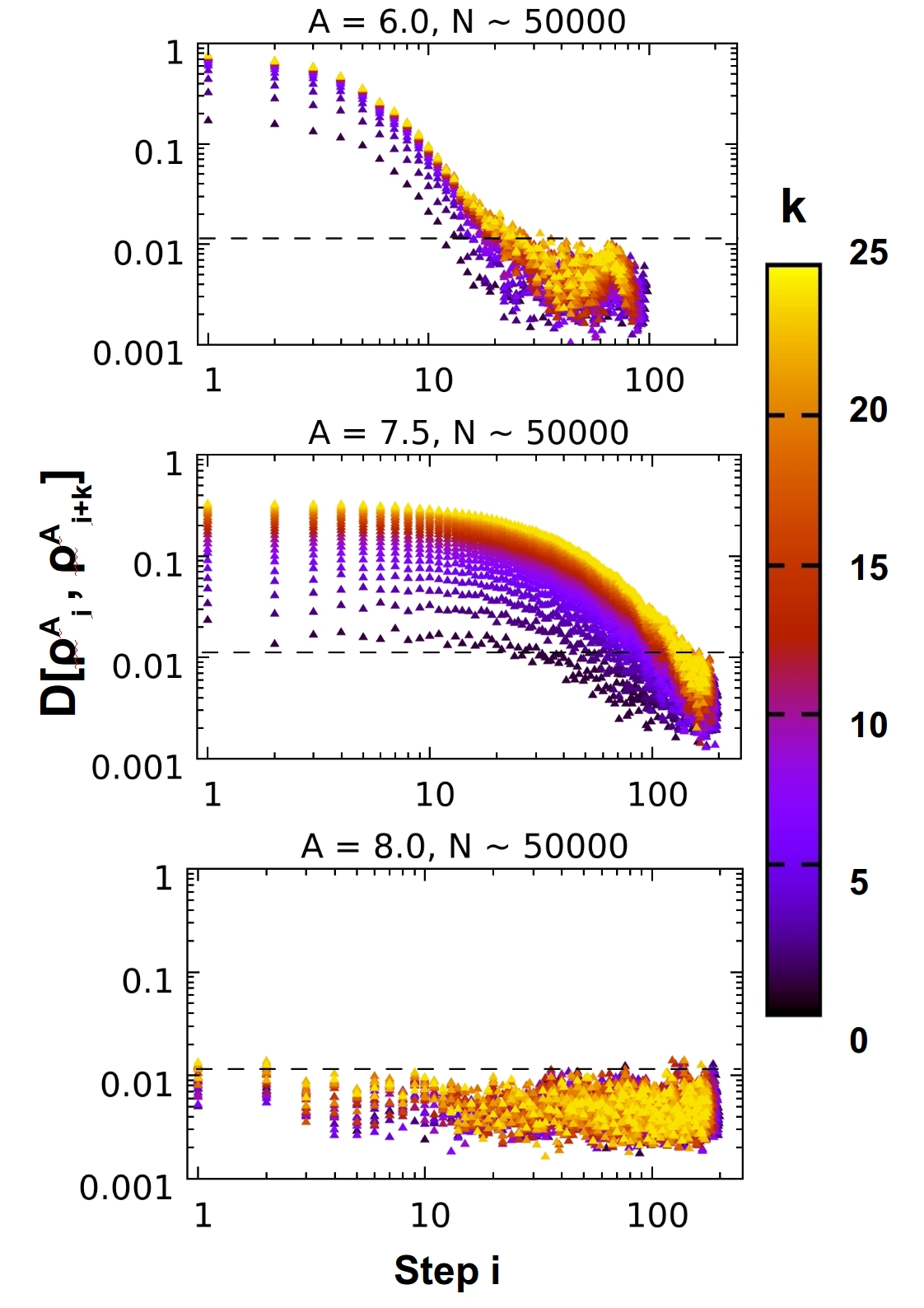}
\caption{Convergence analysis. Plot of the KS distances $D[\rho^{^A}_i,\rho^{^A}_{i+k}]$ ($k$
  running from 1 to 25 being represented by the different colors) for
  different typical amplitudes corresponding to different relaxation
  regimes: $A=6.0$ (top), $A=7.5$ (middle) and $A=8.0$ (bottom). The dashed
  line in each plot corresponds to the critical value $\epsilon^*$ for
  histograms built from 50000 data points.  }
\label{Fig2}
\end{figure}

\subsection{Convergence analysis}

To quantify the convergence oberved for the volume histograms in
Fig.\ref{Fig1}, we want to assess to what extent a sequence $\{
\rho^{^A}_i(V) \}_{i = 1,2,..}$ becomes independent of $i$ for large $i$. For this purpose, we introduce the 2-sample Kolmogorov-Smirnov
(KS) statistic $D[\rho_1,\rho_2]$ between two one-dimensional histograms $\rho_1$ and $\rho_2$ \cite{NR}:
\begin{equation}
D[\rho_1,\rho_2] \equiv \sup_{V}\{|F_1(V)-F_2(V)|\}, \label{eq1}
\end{equation}
where $F_1$ and $F_2$ are the respective cumulative distribution functions
associated with histograms $\rho_1$ and $\rho_2$. This ``distance'' is
commonly used as a test for the {\it null hypothesis} that the two
histograms $\rho_1$ and $\rho_2$ are different realisations of the same
underlying distribution. This hypothesis can be rejected with 99\% of
certainty if \cite{Smirnov48}:
\begin{equation}
 D[\rho_1,\rho_2] > 1.63 \:\sqrt{\frac{n_1+n_2}{n_1n_2}} \equiv \epsilon^*(n_1,n_2), \label{eq2}
\end{equation}where $n_1$ and $n_2$ are the respective number of samples used to build $\rho_1$ and $\rho_2$. From Eq.\eqref{eq2}, we expect that the KS statistic will decrease as the accuracy of each histogram increases (i.e., $\epsilon^*(n_1, n_2)$ decreases), unless the two histograms sample different underlying distributions.

In practice, we seek the existence of an equilibrium step, denoted by
$I(A)$, past which the deviation from the invariant histogram is within the
statistical noise.  In Fig.\ref{Fig2}, we plot the KS distances
$D[\rho^{^A}_i,\rho^{^A}_{i+k}]$ as a function of both $i$ (x-axis) and $k$
(color code). We also display the critical value $\epsilon^* = 0.0109$
corresponding to histograms built from 50000 samples (dashed line,
Eq.\eqref{eq2}). We identify $I(A)$ as the first step for which all colored
points (from black for $k=1$ to bright yellow for $k=25$) are below the
dashed line. We emphasize that it is not enough for
$D[\rho^{^A}_i,\rho^{^A}_{i+1}]$ to be less than $\epsilon^*$. For example,
as shown in Fig.\ref{Fig2}, in the case of $A = 7.5$, this weaker condition
is satisfied (black points) as early as the 30th tap, whereas our
convergence criterion is fulfilled only after the step $I(7.5) = 164$.

We also notice that the relaxation time increases when going from $A=6.0$ to
$A=7.5$, whereas it decreases when going from $A=7.5$ to $A=8.0$.

\begin{figure}[h!]
    \includegraphics[scale=0.22,keepaspectratio=true]{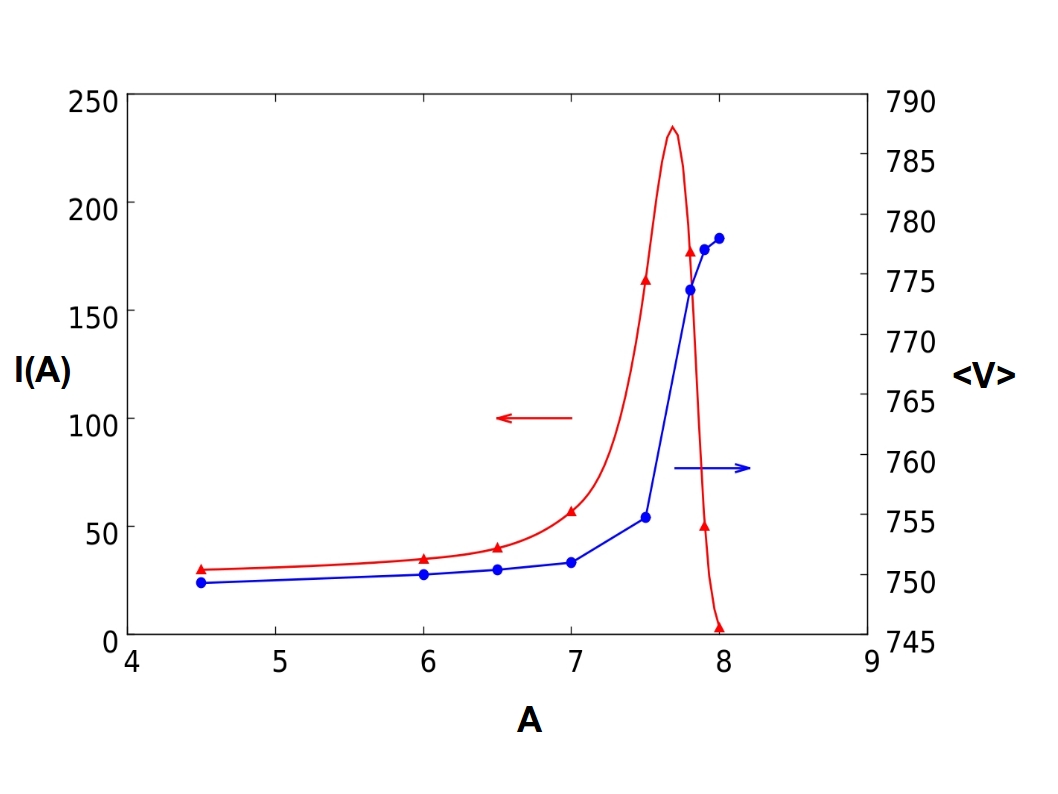}
\caption{Invariant properties. Red curve (triangles): relaxation number of
  steps $I(A)$ vs tapping amplitude, $A$.  The line is a cubic spline
  interpolation, and is meant to guide the eye. Blue curve (circles):
  invariant average volume, $\langle V \rangle$, vs $A$.}
\label{Fig3}
\end{figure}

In Fig.\ref{Fig3}, we plot the mean invariant volumes, $\langle V\rangle$,
and their corresponding equilibrium step, $I(A)$, for the amplitudes we have
tested. We notice that $\langle V\rangle$ increases with tapping amplitude, as found in previous studies \cite{Nowak98,McNamara09,Brey01,Mueggenburg12,Pugnaloni08}. The values
of $I(A)$ are consistent with the number of steps required for the ensemble
average volume to reach a plateau value for a given amplitude $A$ (data not
shown). As stressed previously, we note that it is not an always decreasing function of $A$. To our knowledge, this is the first time that a nonmonotonic
dependence of relaxation time scales on tapping amplitude has been
observed. Normally, one would expect a decrease of the relaxation time as
the tapping amplitude is increased
\cite{Nowak98,McNamara09,Brey01,Mueggenburg12,Pugnaloni08}, although a
nonmonotonic dependence on $A$ was already suggested in Ref.\cite{Brey01}
but in a different context. Our current understanding of this dependence is
the following. At sufficiently low $A$, the ensemble behaves like a quenched
glass.  Within this regime, higher amplitudes result in the system
exploring a larger local basin, which increases the time to sample the basin
and, hence, the relaxation time. On the other hand, if we were to tap on the
system strongly enough so that each shaking step is akin to our preparation
protocol, then we should not observe any change in the ensemble histogram
between successive taps, which would yield a vanishing relaxation time.
These two limits can only be reconciled if the relaxation time depends
nonmonotonically on tapping amplitude, as shown in
Figs. \ref{Fig2} and \ref{Fig3}.

\section{Tests on the invariant histograms}
\subsection{Ergodicity analysis}
\begin{figure}[h!]
    \includegraphics[scale=0.235,keepaspectratio=true]{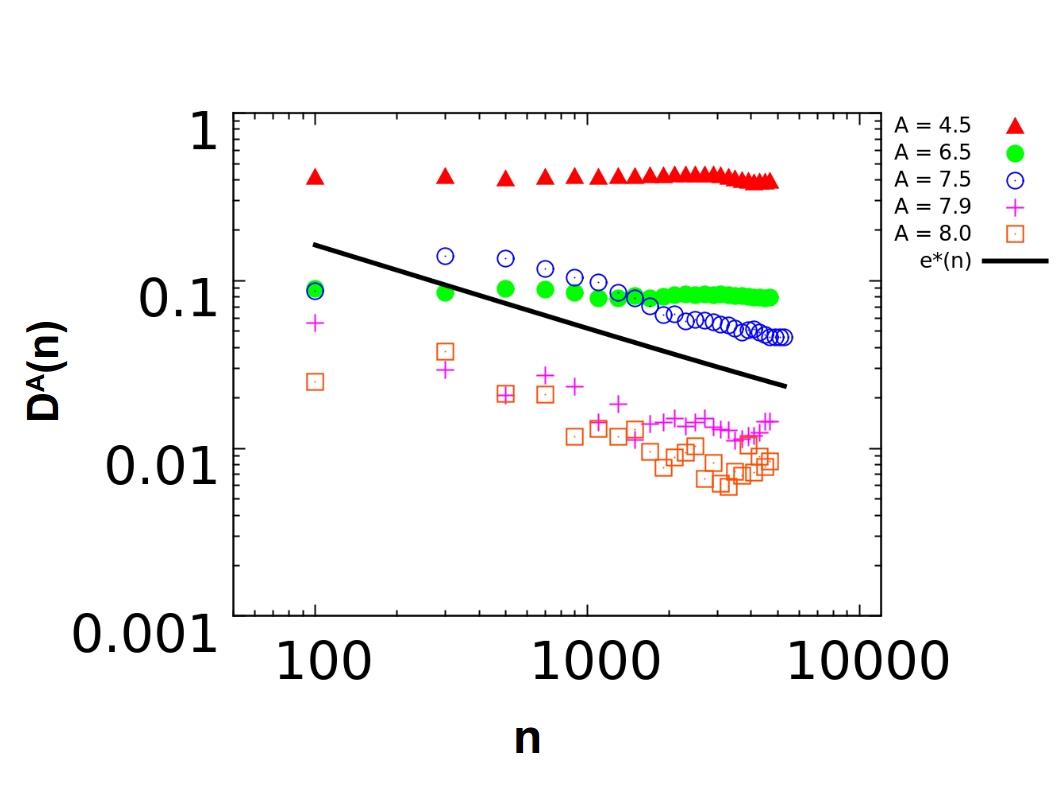}
\caption{Ergodicity analysis. Plot of the distance $D^{^A}(n)$ as a function
  of the number of uncorrelated volume values, $n$, for different amplitudes, $A$. The plain black
  line corresponds to the threshold value $\epsilon^*(n)$ calculated from
  Eq.\eqref{eq2}. The sampling is done after having tapped $200$ times on
  the system.  }
\label{Fig4}
\end{figure}

To explain the behaviour observed in
Figs. \ref{Fig2} and \ref{Fig3}, we have proposed in the preceding section that some conditions lead
to nonergodicity. In this section, we test this proposal quantitatively.

Denote by $\rho^{^A}_n(V)$ a volume histogram built from $n$ uncorrelated volume values taken from a very long trajectory of one system tapped with an amplitude $A$. In practice, we consider a single trajectory for a system tapped about $10^5$ times and we compute the corresponding volume correlation function. We then use the latter to select $n$ uncorrelated values belonging to this trajectory.  Under the ergodic hypothesis, this
``time'' histogram and the invariant ensemble histogram, $\rho^{^A}_{i >
  I(A)}(V)$, both sample the same underlying distribution. To test this
idea, we again use the KS statistic.  Let $D^{^A}(n)$ be the quantity
$D[\rho^{^A}_n(V), \rho^{^A}_{i > I(A)}(V)]$. If $D^{^A}(n)$ is bigger than
the rejection value $\epsilon^*(n)$ calculated from Eq.\eqref{eq2}, with
$n_1 = n$ and $n_2 = 50000$, then the ergodic hypothesis is rejected.
Recall that the KS statistic decreases with increasing $n$ if and only if
the {\it null hypothesis}, ergodicity, is true. Otherwise, $D^{^A}(n)$
saturates at a finite value when $n$ is large enough. Fig.\ref{Fig4} shows
that for high values of $n$, $D^{^A}(n)$ indeed tends to saturate. In that case, the difference between the saturation value of $D^{^A}(n)$ and the curve $\epsilon^*(n)$ at a fixed $n$ serves as a useful measure of
nonergodicity that we will call the {\it ergodicity gap}.
Fig.\ref{Fig4} shows that for tapping amplitudes $A = 4.5,\:6.5$ and $7.5$, trajectories are nonergodic, with ergodicty gaps that increase with decreasing tapping amplitude. In contrast, for $A = 7.9$ and $8.0$, trajectories are
ergodic according to our criterion. These findings are consistent with our
current understanding of the nonmonotonic trend for $I(A)$ observed in
Figs. \ref{Fig2} and \ref{Fig3}.

\subsection{Compatibility with Edwards' prior}
In 1989, Edwards et al. \cite{Edwards89} suggested that if a granular
material is submitted to a protocol that yields different packing volumes
then the probability of occurence of a given stable configuration is
proportional to a Boltzmann weight $e^{-\beta V}$, where $V$ is the actual
volume of the packing configuration. Since then, many studies have tested
this idea and its consequences, but no consensus has been reached
\cite{Blumenfeld09, Henkes09,Bowles11, Aste08, Wang10,Lechenault10, Aste08,
  McNamara09, Nowak98}.

In their study, McNamara et al. tested Edwards' prior by generating
time-sampled volume histograms for different tapping amplitudes and looking
directly at ratios between between pairs of histograms.  This procedure is
known as the {\it overlapping histogram method} \cite{Bennett76}: if the
logarithm of the ratio between two histograms is linear, then there is a
potential compatibility with Edwards' canonical assumption.  Otherwise, this
assumption should be rejected, at least for the tested protocol. McNamara et
al. found a linear behaviour for the log ratios, for the amplitudes that
they tested both numerically and experimentally. Here, we apply the same
test on our own invariant ensemble histograms.

\begin{figure}[h!]
    \includegraphics[scale=0.235,keepaspectratio=true]{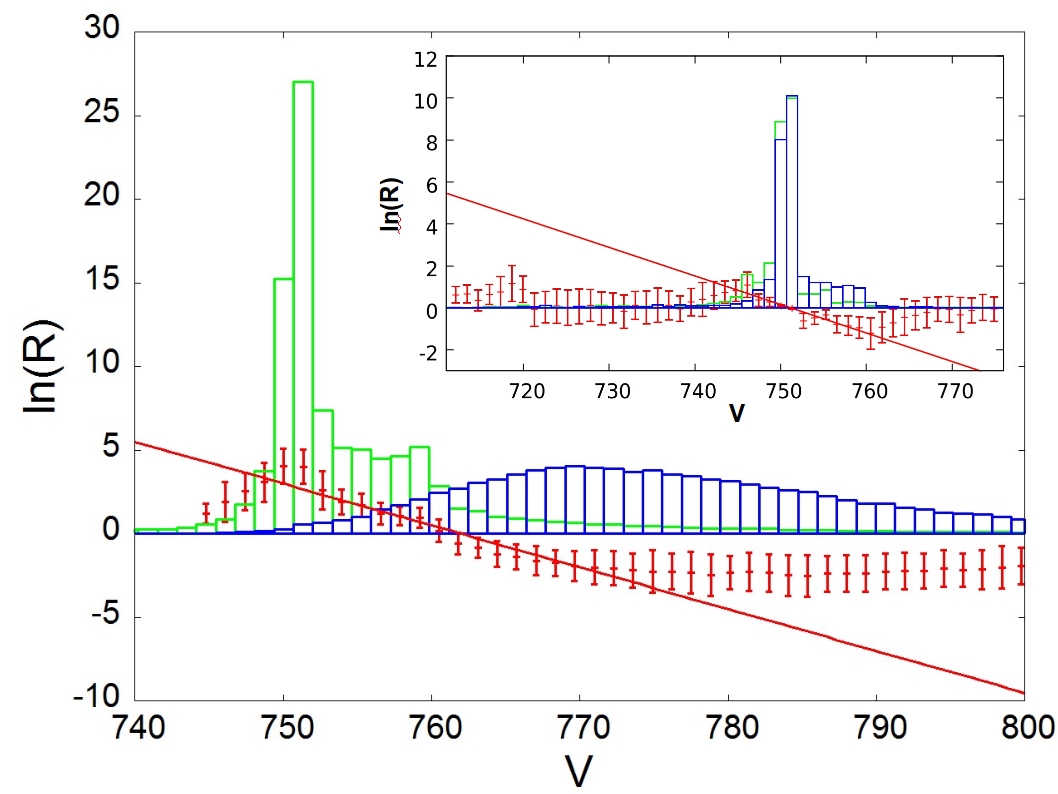}
\caption{Overlapping Histograms. The red points correspond to the plot of the logarithm of the ratios $\rho^{7.5}_{i > I_{7.5}}/\rho^{8.0}_{i > I_{8.0}}$ (main figure) and $\rho^{4.5}_{i > I_{4.5}}/\rho^{6.5}_{i > I_{6.5}}$ (inset). We superimposed the corresponding invariant histograms in both the main figure and the inset. The histogram in green always corresponds to the smallest amplitude in a given pair. The plain lines are the best linear fits through the midpoints within the crossover regions.
}
\label{Fig5}
\end{figure}

Fig.\ref{Fig5} shows the log ratio of two invariant ensemble histograms ($A
= 7.5$ with $A = 8.0$ for the main plot and $A=4.5$ with $A=6.5$ in the inset). We find that over a limited range of volumes where
the histograms overlap significantly, there is indeed a linear decrease in
the log ratio of the two invariant histograms, consistent with the findings
in Ref.\cite{McNamara09}. However, we also find that for high volumes, this
ratio tends to saturate, whereas for low volumes, it rapidly decreases. These trends are present for all pairs of invariant volume
histograms that we have tested, including those where the two histograms
are apparently ``closeby''.

It is worth noting that the authors of Ref.\cite{McNamara09} have already
suggested the possibility that the observed linearity is only true locally.
Moreover a close look at their own figures reveals deviations from linearity
for very high and very low volumes that are consistent with our own
observations (Fig.\ref{Fig5}).

Overall,  Edwards' canonical hypothesis as a global property is not compatible with the protocol we are testing. Although this is not the first
time that a strong disagreement with Edwards' theory has been found
\cite{Gao06}, to our knowledge, this is the first time that a full
histogram analysis reports incompatibility with Edwards' canonical
assumption for vibrated granular matter. Our study does not rule out the possibility of a local compatibility with Edwards' assumption but this is already different from Edwards' orignal theory.

\section{Conclusion}
In this letter, we introduced an ensemble volume satistic for a simulated vertically vibrated granular system. To quantitatively assess the properties of the generated sequences of ensemble histograms, we used the KS test. This allowed us to devise a convergence criterion for a sequence of histograms. Subsequently, we tested the ergodicity of a tapped system as a function of the tapping amplitude, $A$, and found clear evidence for nonergodicity when the tapping amplitude is low. Finally, we tested the compatibility of our invariant histograms with Edwards' hypothesis and concluded that it is not compatible with our simulation protocol.

We should point out that the results found in this paper depend {\it a priori} on the chosen tapping protocol and also on the preparation stage. Although the dependence of our findings on the preparation protocol is hard to predict, the tools that we introduced to characterize our ensemble histograms can be used to test any numerical or experimental protocol.

\acknowledgments{Fabien Paillusson is grateful to Nicolas Dorsaz, Frank Smallenburg  and Patrick Varilly for very helpful discussions. This work has been supported by the EPSRC grant $N^{\circ}$ EP/I000844/1. D.F. acknowledges support from ERC Advanced Grant 227758, Wolfson Merit Award 2007/R3 of the Royal Society of London.}
\bibliographystyle{apsrev4-1}
\bibliography{biblio_granular}
\end{document}